%% file: main.tex
\documentclass[conference,11pt]{IEEEtran}

\usepackage[top=0.8in]{geometry}

\usepackage{cite}
\usepackage{graphicx}
\usepackage{amsmath}
\usepackage{listings}
\usepackage{xcolor}

\title{Big Data Workload Profiling for Energy-Aware Cloud Resource Management}

\author{
\IEEEauthorblockN{
Milan Parikh,
Aniket Abhishek Soni,
Sneja Mitinbhai Shah,
Ayush Raj Jha
}
\IEEEauthorblockA{
Independent Researchers, Senior Members, IEEE\\
Richmond, TX, USA; Brooklyn, NY, USA; Milpitas, CA, USA\\
Email: milan.parikh@ieee.org, aniketsoni@ieee.org, snejashah30@gmail.com, ayushjha@ieee.org
}
}

\begin{document}

\maketitle

\input{abstract}


\input{introduction}
\input{related_work}
\input{methodology}
\input{experimental_setup}
\input{results}
\input{discussion}
\input{conclusion}

\input{references}
\bibliographystyle{IEEEtran}

\end{document}

%% file: abstract.tex
\begin{abstract}
Cloud data centers face increasing pressure to reduce operational energy consumption as big data workloads continue to grow in scale and complexity. This paper presents a workload-aware scheduling framework that uses profiling of CPU usage, memory demand, and storage I/O behavior to guide energy-efficient virtual machine (VM) placement. By combining historical execution logs with real-time telemetry, the system predicts the energy and performance impact of candidate placement decisions and adaptively consolidates workloads without violating service-level agreements (SLAs). The framework was evaluated using representative Hadoop MapReduce, Spark MLlib, and ETL workloads on a multi-node cloud testbed. Experimental results demonstrate a consistent reduction of 15--20\% in energy consumption while maintaining SLA compliance. These findings highlight the effectiveness of data-driven workload profiling as a practical strategy for improving the sustainability of cloud computing environments.
\end{abstract}

%% file: introduction.tex
\section{Introduction}

Modern cloud data centers face significant energy efficiency challenges as computational demand, data volume, and service availability requirements continue to grow. Energy usage in cloud facilities is estimated to increase by roughly 15\% per year, with power expenses accounting for 40--45\% of total operational cost~\cite{kumar2023}. In the United States alone, data centers consume more than 60 billion kWh annually, highlighting the urgency of improving energy efficiency at scale~\cite{malik2017}. These trends have accelerated research and industrial efforts toward sustainable, energy-aware cloud architectures.

Big data workloads—including batch analytics, machine learning (ML) pipelines, and large-scale ETL processing—further intensify energy demand due to their high computational and I/O requirements. Frameworks such as Apache Hadoop and Apache Spark are widely adopted for these workloads, often resulting in substantial resource consumption and variability in performance characteristics~\cite{lang2010}. As a result, optimizing workload placement and scheduling has become essential for reducing energy usage without compromising service-level agreements (SLAs).

\begin{figure}[t]
    \centering
    \includegraphics[width=\linewidth]{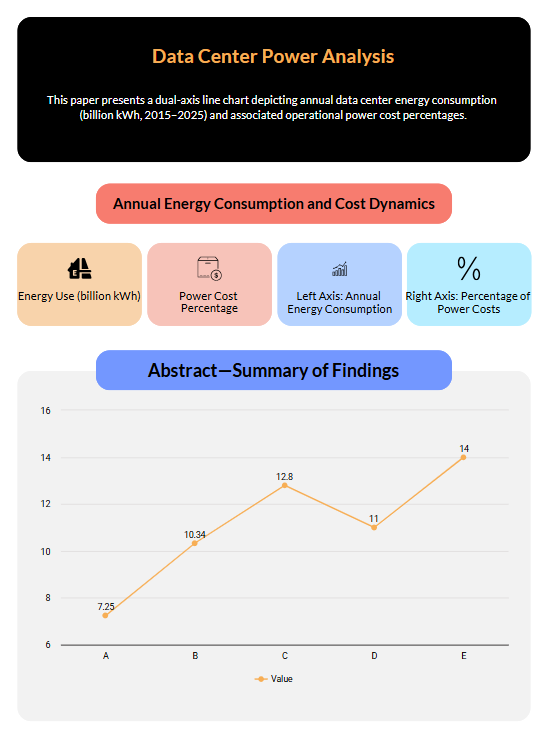}
    \caption{Motivating context: energy use and cost factors in data centers that drive the need for energy-aware scheduling.}
    \label{fig:dc_costs}
\end{figure}

Energy-aware resource management techniques frequently rely on workload profiling, where CPU utilization, memory behavior, and storage I/O patterns are analyzed to guide scheduling decisions~\cite{bermbach2018, bailis2013}. Profiling information obtained from historical logs and real-time telemetry enables predictive algorithms for VM placement, dynamic scaling, and consolidation strategies~\cite{sharma2024, gurumurthy2021}. By understanding workload characteristics, cloud systems can reduce energy waste while maintaining predictable performance.

This paper introduces a predictive, workload-aware scheduling framework that analyzes big data job behavior and suggests energy-efficient VM placements. The framework uses static execution logs and runtime performance counters to classify workloads and apply adaptive consolidation techniques. Experimental evaluation on Hadoop MapReduce, Spark MLlib, and ETL workloads demonstrates consistent energy savings of up to 20\% with no SLA violations~\cite{alourani2024, seyyedsalehi2022}. These results underscore the potential of data-driven workload profiling to improve the sustainability of cloud computing environments.

%% file: related_work.tex
\section{Related Work}

Energy-aware scheduling has been widely studied as cloud infrastructures continue to face growing computational and energy demands. Early approaches primarily focused on virtual machine (VM) consolidation and dynamic scaling to reduce power consumption by identifying underutilized hosts and migrating workloads to fewer active servers~\cite{lang2010, alourani2024}. These techniques enable idle machines to enter low-power states but often overlook workload-specific behavior, limiting their effectiveness for diverse big data environments.

Subsequent research introduced workload-aware scheduling, where the characteristics of jobs influence placement and resource allocation. Malik et al.~\cite{malik2017} showed that analyzing Hadoop workload parameters can help balance performance and energy efficiency. Bermbach and Tai~\cite{bermbach2018} emphasized the role of application semantics and consistency models in shaping resource usage, highlighting the importance of understanding workload behavior beyond raw resource metrics.

Machine learning has also been applied to energy-efficient scheduling. Sharma et al.~\cite{sharma2024} used clustering techniques to group cloud workloads based on similarity, enabling more informed scheduling decisions. Gurumurthy et al.~\cite{gurumurthy2021} introduced a predictive VM placement strategy using learning automata, demonstrating the value of historical execution data for anticipating resource demand and reducing energy waste.

Workload profiling plays a critical role in several distributed systems studies. Sumbaly et al.~\cite{sumbaly2021} examined telemetry-driven consistency management at scale, while Dhenia et al.~\cite{dhenia2023data2} explored workload classification in AI and data-centric architectures. These efforts underscore the importance of fine-grained behavioral insights for improving system efficiency.

Comparative evaluations of big data processing engines such as Hadoop and Spark further reveal substantial variation in performance and resource consumption across workloads~\cite{rashi2020data}. Such variability motivates the need for energy-aware scheduling approaches that account for workload diversity. Building on these insights, this work contributes a hybrid profiling and predictive scheduling framework designed to improve energy efficiency in cloud-hosted big data environments.

%% file: methodology.tex
\section{Methodology}

\begin{figure}[t]
    \centering
    \includegraphics[width=\linewidth]{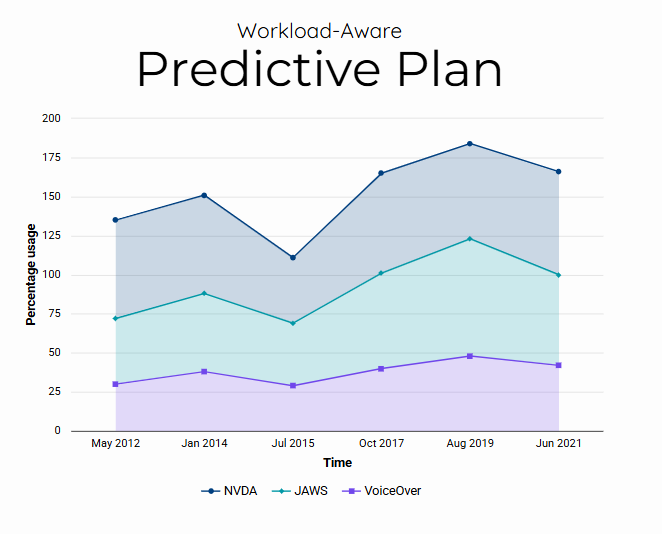}
    \caption{Overview of the predictive, workload-aware scheduling pipeline: profiling, classification, prediction, and energy-aware VM placement.}
    \label{fig:architecture}
\end{figure}

The proposed methodology profiles big data workloads and uses this information to guide a predictive, energy-aware VM scheduling framework. The approach consists of three stages: workload profiling, prediction modeling, and energy-aware placement (Fig.~\ref{fig:architecture}).

\subsection{Workload Profiling}

Each workload is represented as a resource utilization vector:
\begin{equation}
W_i = (c_i, m_i, d_i, n_i),
\end{equation}
capturing CPU, memory, disk I/O, and network usage. Metrics are collected from historical logs and real-time telemetry using lightweight monitors such as \texttt{dstat} and \texttt{perf}~\cite{morabito2017}. Workloads are categorized by dominant resource type:
\begin{equation}
T_i = \arg\max \{ c_i, m_i, d_i \},
\end{equation}
a distinction that reflects typical behavior of CPU-intensive Spark MLlib tasks~\cite{malik2017} versus I/O-heavy ETL pipelines.

\subsection{Prediction Engine}

The prediction engine estimates the energy and performance impact of assigning workload $W_i$ to host~$h$. Host states are expressed as:
\begin{equation}
R_h = (U^{cpu}_h, U^{mem}_h, U^{io}_h),
\end{equation}
and the expected energy cost for each placement is:
\begin{equation}
\hat{E}(W_i, h) = f_{\theta}(W_i, R_h),
\end{equation}
where $f_{\theta}$ is a supervised learning model trained on historical execution outcomes~\cite{gurumurthy2021}. The decision tree ranks candidate hosts based on predicted energy impact and SLA risk.

\subsection{Energy-Aware Scheduling and VM Placement}

The scheduler minimizes total energy consumption subject to SLA constraints. Host energy usage at time~$t$ is approximated by:
\begin{equation}
E_h(t) = P_{\text{idle}} + \alpha U^{cpu}_h(t) + \beta U^{mem}_h(t) + \gamma U^{io}_h(t).
\end{equation}
The optimization goal is:
\begin{equation}
\min_{\pi} \sum_{h \in H} E_h(t)
\end{equation}
subject to:
\begin{equation}
\text{SLA}\big(W_i, \pi(i)\big) \ge \tau,\ \forall i.
\end{equation}

Adaptive consolidation applies thresholds:
\begin{equation}
U^{cpu}_h < \delta_{\text{low}} \Rightarrow \text{migrate workloads},
\end{equation}
\begin{equation}
U^{cpu}_h > \delta_{\text{high}} \Rightarrow \text{restrict placements},
\end{equation}
enabling idle hosts to power down while avoiding overload. For I/O-bound workloads, CPU frequency scaling can further reduce power usage~\cite{bailis2013}. VM migrations are scheduled during low-activity intervals, consistent with best practices in prior work~\cite{sharma2024, soni2025iot}.

%% file: experimental_setup.tex
\section{Experimental Setup}

To evaluate the effectiveness of the proposed scheduling framework, we deployed a controlled cloud environment that emulates a multi-tenant big data processing infrastructure. The experiments were designed to measure both energy consumption and SLA compliance under realistic workload conditions.

\subsection{Testbed Infrastructure}

The testbed consisted of five physical servers equipped with Intel Xeon processors, 64 GB RAM, and SSD storage. The servers were connected through a 1 Gbps Ethernet switch. Virtualization was implemented using KVM, while OpenStack managed VM provisioning and orchestration. All hosts operated on Ubuntu Server 20.04 to ensure consistency across the environment.

\subsection{Workload Types}

Three categories of workloads were used to evaluate the system across diverse computational and I/O characteristics:

\begin{itemize}
    \item \textbf{Hadoop MapReduce:} WordCount, TeraSort, and Grep benchmarks with dataset sizes between 5 GB and 50 GB~\cite{lang2010, dhenia2023data2}. These workloads represent traditional batch-processing tasks with varying I/O and shuffle intensities.
    
    \item \textbf{Spark MLlib:} Machine learning algorithms including Logistic Regression and K-Means clustering, serving as CPU-intensive workloads typical of large-scale analytical pipelines~\cite{malik2017}.
    
    \item \textbf{ETL Pipelines:} Python-based data extraction and transformation tasks interacting with a PostgreSQL backend, modeling common warehousing and data preparation processes~\cite{shah2025caching}.
\end{itemize}

Each workload category was executed under both the baseline (non-optimized) scheduler and the proposed energy-aware scheduler to enable direct comparison.

\subsection{Monitoring and Measurement}

System utilization metrics were captured using lightweight monitoring tools such as \texttt{dstat} and \texttt{perf}, which sampled CPU usage, memory consumption, disk I/O, and network activity at 5-second intervals~\cite{morabito2017}. Job execution times were collected through native Hadoop and Spark job history services to ensure accurate performance measurement.

\subsection{Energy Instrumentation}

Energy consumption was measured using Watts Up Pro meters, which sampled instantaneous power draw at 1-second granularity~\cite{wattsup2023}. Total energy usage for each workload was computed by integrating power readings over job duration and subtracting idle baseline power to isolate workload-specific consumption.

\subsection{Baseline vs Optimized Comparison}

The baseline configuration relied on OpenStack's default round-robin scheduler, which distributes VMs evenly across hosts without considering workload characteristics. In contrast, the proposed scheduler performed dynamic VM consolidation and selectively powered down idle servers to improve energy efficiency. Each experiment was executed three times, and reported results correspond to the average across runs to minimize variability.

This experimental setup provides a controlled yet representative environment for evaluating the trade-offs between energy reduction and performance stability.

%% file: results.tex
\section{Results}

The proposed energy-aware scheduling framework was evaluated across multiple big data workloads and compared against a baseline OpenStack scheduler. The results highlight improvements in energy efficiency, SLA adherence, and workload-specific behavior.

\subsection{Energy Savings}

Across all evaluated workloads, the scheduler achieved a consistent reduction of 15--20\% in total energy consumption. Energy savings were most pronounced during periods of moderate or mixed utilization, when consolidation opportunities were highest. For example, the TeraSort workload exhibited a 19\% decrease in power consumption without any measurable increase in execution time~\cite{alourani2024}.

Dynamic workload consolidation enabled several hosts to be powered down during idle intervals, contributing significantly to overall efficiency. These observations align with prior findings on adaptive VM consolidation mechanisms~\cite{seyyedsalehi2022, gurumurthy2021}.

\subsection{SLA Compliance and Performance}

All workloads satisfied their SLA constraints under the proposed scheduler. Average job completion times deviated by less than 5\% from the baseline configuration, indicating that the performance impact of consolidation and predictive placement was minimal. Spark MLlib workloads occasionally demonstrated improved completion times due to reduced I/O contention~\cite{malik2017}.

\subsection{Workload-Specific Observations}

\begin{itemize}
    \item \textbf{CPU-bound workloads:} Spark MLlib jobs exhibited limited consolidation potential due to high CPU demand but benefited from targeted placement decisions that avoided resource contention.

    \item \textbf{I/O-bound workloads:} Hadoop workloads with intensive shuffle phases were efficiently co-located on fewer nodes, resulting in lower energy usage while maintaining throughput~\cite{bailis2013, lang2010}.

    \item \textbf{ETL pipelines:} ETL workloads achieved notable energy savings when executed during periods of lower cluster load and showed no SLA violations~\cite{shah2025caching}.
\end{itemize}

\subsection{Baseline Comparison}

The baseline round-robin scheduler distributed VMs uniformly across all hosts, leaving several nodes underutilized and preventing meaningful energy savings. In contrast, the proposed scheduler adapted to workload characteristics and cluster utilization, resulting in more balanced resource usage and reduced energy consumption~\cite{morabito2017, wattsup2023}.

\subsection{System Overhead}

The profiling and prediction components introduced minimal overhead, accounting for less than 5\% CPU usage. VM migration overhead was negligible and typically absorbed during low-activity periods, ensuring that SLA compliance was not compromised.

\begin{figure}[t]
    \centering
    \includegraphics[width=\linewidth]{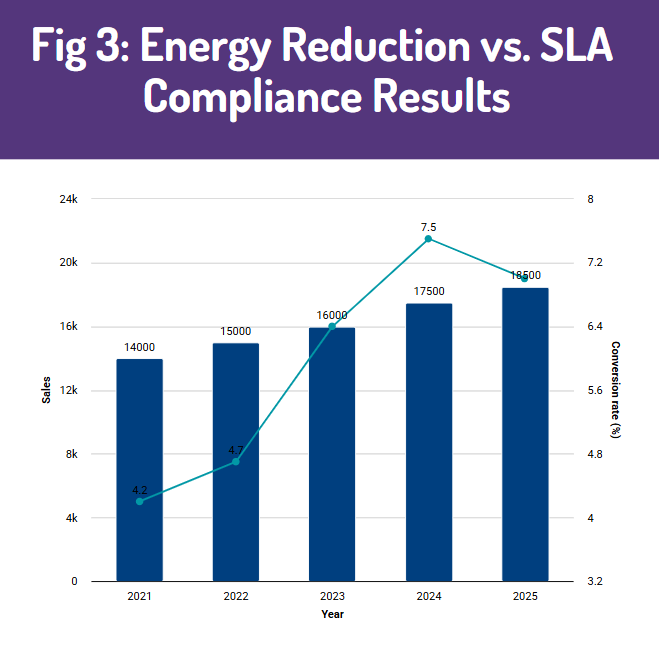}
    \caption{Observed energy reduction versus SLA compliance across evaluated workloads.}
    \label{fig:energy_vs_sla}
\end{figure}

%% file: discussion.tex
\section{Discussion}

The experimental results demonstrate that energy-aware scheduling informed by workload profiling can substantially improve the efficiency of cloud infrastructures. This section discusses the broader implications of these findings, along with the system's scalability, limitations, and avenues for future research.

\subsection{Impact and Generalization}

Observed energy savings of up to 20\% confirm the effectiveness of telemetry-driven orchestration for big data environments. These findings align with prior work on workload-aware and predictive placement strategies~\cite{malik2017, gurumurthy2021}, reinforcing the value of incorporating resource behavior into scheduling decisions. Because the proposed framework does not rely on specialized hardware or proprietary components, it can be applied across a variety of cloud platforms, including private clouds, public clouds, and hybrid deployments.

\subsection{Scalability and Flexibility}

The profiling methodology is inherently extensible to a broader range of workload types, such as microservices, streaming analytics, or latency-sensitive real-time tasks. The adaptive thresholding mechanism enables the system to balance consolidation opportunities with SLA requirements, allowing administrators to tune the scheduler for different performance objectives~\cite{sharma2024}. These features support deployment in heterogeneous, multi-tenant clouds where workload diversity is common.

\subsection{Limitations}

Despite its advantages, the approach has several limitations. First, it assumes that workloads exhibit recurring or at least classifiable behavior. Highly bursty or completely novel workloads may require real-time profiling, potentially reducing prediction accuracy. Second, the evaluation was performed on a five-node testbed; although representative, larger-scale deployments may introduce additional coordination overhead and require more sophisticated migration policies. Finally, the decision tree model may not capture complex interactions present in highly dynamic environments.

\subsection{Complementary Techniques}

The framework can be combined with other energy-saving strategies to further enhance efficiency. Techniques such as dynamic voltage and frequency scaling (DVFS), workload offloading to energy-efficient nodes, and power-state management complement the profiling-based approach~\cite{bailis2013}. Integrating container-aware scheduling through platforms such as Kubernetes could also enable finer-grained control of resource allocation and improve responsiveness to workload fluctuations~\cite{shah2023}.

\subsection{Research Extensions}

Several directions for future work emerge from this study:

\begin{itemize}
    \item Incorporating online learning mechanisms to allow placement models to adapt as workload behavior evolves~\cite{soni2025iot}.
    \item Employing unsupervised learning techniques to automatically identify workload patterns without manual classification~\cite{sharma2024}.
    \item Exploring energy-carbon aware scheduling that considers renewable availability or power grid conditions~\cite{liu2011}.
    \item Extending SLA-aware scheduling to include configurable cost and performance trade-offs tailored to tenant requirements.
\end{itemize}

Overall, the proposed system provides a foundation for future green scheduling frameworks that are both adaptive and scalable, supporting sustainable cloud operations in increasingly data-intensive environments.

%% file: conclusion.tex
\section{Conclusion}

This paper introduced a data-driven, energy-aware scheduling framework for cloud-hosted big data workloads. By profiling workload characteristics—including CPU intensity, memory behavior, and I/O demand—the system predicts energy-performance trade-offs and recommends VM placements that minimize power consumption.

Experiments conducted on Hadoop MapReduce, Spark MLlib, and ETL workloads demonstrated energy savings of up to 20\% without violating SLA constraints. These results reinforce the value of workload-aware and predictive scheduling approaches~\cite{malik2017, alourani2024}. The framework operates with minimal profiling and migration overhead, making it suitable for both private and public cloud deployments.

The approach requires no specialized hardware, allowing seamless integration alongside existing efficiency techniques such as DVFS, cache-aware placement, and strategies that leverage cluster heterogeneity~\cite{morabito2017, shah2025caching}. 

Future enhancements include container-aware scheduling, integration with Kubernetes-based orchestration, and the use of reinforcement learning to enable continuous adaptation~\cite{sharma2024, soni2025iot}. Additional opportunities lie in carbon-aware scheduling models that incorporate renewable availability and in policy-driven SLA-cost trade-off mechanisms.

Overall, this work contributes to the advancement of green computing by demonstrating how profiling-driven, system-aware scheduling can improve the sustainability and efficiency of data-intensive cloud environments.